\documentclass[prd,aps,preprintnumbers, showpacs, nofootinbib,superscriptaddress,notitlepage]{revtex4-1}

\usepackage{amsmath,graphicx,epsfig,amssymb,dsfont,mathtools}
\usepackage[usenames]{color}
\usepackage{ulem} %% for strike-through
\usepackage{bigstrut}
\usepackage{slashed}
\usepackage{multirow}
\usepackage{subfigure}

\allowdisplaybreaks

\begin{document}
	
		\preprint{IBS-CTPU-23-57}
	
\title{Light baryon in three quark picture light front approach and its application:\\ hyperon weak radiative decays}

\author{Zhi-Peng Xing}
\email{zpxing@nnu.edu.cn}
\affiliation{Department of Physics and Institute of Theoretical Physics, Nanjing Normal University, Nanjing, Jiangsu 210023, China}
	
\author{Yu-Ji Shi}
\email{shiyuji@ecust.edu.cn}
\affiliation{School of Physics, East China University of Science and Technology, Shanghai 200237, China }

\author{Jin Sun}
\email{sunjin0810@ibs.re.kr(Corresponding author)}
\affiliation{Particle Theory and Cosmology Group, Center for Theoretical Physics of the Universe, Institute for Basic Science (IBS), Daejeon 34126, Korea}

%\author{Wei Wang}
%\email{wei.wang@sjtu.edu.cn}
%\affiliation{INPAC, Key Laboratory for Particle Astrophysics and Cosmology (MOE), Shanghai Key Laboratory for Particle Physics and Cosmology,
%School of Physics and Astronomy, Shanghai Jiao Tong University, Shanghai 200240, China}
%\affiliation{Southern Center for Nuclear-Science Theory (SCNT), Institute of Modern Physics, Chinese Academy of Sciences, Huizhou 516000, Guangdong Province, China}

\author{ Zhen-Xing Zhao}
\email{zhaozx19@imu.edu.cn}
\affiliation{School of Physical Science and Technology, Inner Mongolia University, Hohhot 010021, China}

\begin{abstract}
Motivated by recent experimental data on $\Sigma^+\to p\gamma$ at BESIII, we investigate a class of hyperon weak radiative decays.
To estimate these processes, in our research, we employ a new type of light-front quark model with a three-quark picture for octet baryons.
In the  three-quark picture, with the use of $SU(3)_f$ and spin symmetries, we present a general form of the light front wave function for each octet baryon.
By including contributions from the penguin diagram and W exchange diagram, we perform a complete calculation on the branching ratios ($Br$) and the asymmetry parameter ($\alpha$) for hyperon weak radiative decay processes.
Our results are helpful for discovering additional hyperon weak radiative decay processes in experimental facilities, and our research will promote the theoretical study of baryons.
\end{abstract}

\maketitle

%%%%%%%%%%%%%%%%%
\section{Introduction}
%%%%%%%%%%%%%%%%

Recently, the BESIII experiment published its precise measurement of the decay $\Sigma^+\to p\gamma$ using $(10087\pm 44)\times 10^{6}$ $J/\psi$ events~\cite{BESIII:2023fhs}. This is the first time this process has been measured in an electron-positron collider experiment. The absolute branching fraction and decay asymmetry parameter have been determined as
\begin{eqnarray}
Br(\Sigma^+\to p\gamma)=(0.996\pm0.021\pm0.018)\times 10^{-3},\;\;\; %\notag\\
\alpha(\Sigma^+\to p\gamma)=-0.652\pm0.056\pm0.020.
\end{eqnarray}
In addition to this process, $\Lambda\to n\gamma$ and other radiative decay was also measured~\cite{Li:2016tlt,Belle:2016mtj,BESIII:2022rgl,Belle:2022raw,BESIII:2022aok,ParticleDataGroup:2022pth,BESIII:2023utd}. To date,  these precise measurements have also promoted theoretical studies of weak radiative decay processes~\cite{Chang:2000hu,Dubovik:2008zz,Zenczykowski:2020hmg,Niu:2020aoz,Wang:2020wxn,Shi:2022dhw,Shi:2022zzh,Adolph:2022ujd,Shi:2023kbu}.

In recent years, there are many theoretical studies focusing on the hyperon~\cite{RQCD:2019hps,Kim:2021zbz,Kim:2021zbz,Irshad:2022zga,Dai:2023vsw,Deng:2023csv,Han:2023xbl,Han:2023hgy}. In previous studies, hyperon weak radiative decays have provided valuable insights into the nature of nonleptonic weak interactions~\cite{Behrends:1958zz}.  Being a two-body decay process, the amplitudes of spin-$\frac{1}{2}$ hyperon radiative decays are described by both a parity-conserving (P-wave) and a parity-violating (S-wave) amplitude.  In Ref.~\cite{Hara:1964zz}, the parity-violating amplitudes were predicted to be zero for $\Sigma\to p\gamma$ and $\Xi\to \Sigma\gamma$ processes in SU(3) symmetry, but this conflicts with the experimental measurement of the asymmetry parameter $\alpha$.  Therefore, it is necessary for us to provide a theoretical analysis of these processes.

For hyperon weak radiative decay processes, such as $\Sigma^+ \to p \gamma$ induced by $s\to d$, the amplitudes are contributed by the W-loop  and the W-exchange diagrams. The corresponding effective Hamiltonians are~\cite{Buchalla:1995vs}
\begin{eqnarray}
&&\mathcal{H}(s\to  d\gamma)=\frac{G_F}{\sqrt{2}}V_{us}V^*_{ud}\sum_i^{1-6,12}[(z_i(\mu)+\tau y_i(\mu))Q_i(\mu)],\nonumber\\
&&Q_1= [\bar u_\alpha\gamma_\mu(1-\gamma_5)s_\beta][\bar d_\beta\gamma^\mu(1-\gamma_5)u_\alpha],\;\;%\nonumber\\
 Q_2=[\bar u_\alpha\gamma_\mu(1-\gamma_5)s_\alpha][\bar d_\beta\gamma^\mu(1-\gamma_5)u_\beta],\notag\\
&&Q_3= [\bar d_\alpha\gamma_\mu(1-\gamma_5)s_\alpha]\sum_q[\bar q_\beta\gamma^\mu(1-\gamma_5)q_\beta],\;\; %\nonumber\\
Q_4= [\bar d_\alpha\gamma_\mu(1-\gamma_5)s_\beta]\sum_q[\bar q_\beta\gamma^\mu(1-\gamma_5)q_\alpha],\nonumber\\
&&Q_5= [\bar d_\alpha\gamma_\mu(1-\gamma_5)s_\alpha]\sum_q[\bar q_\beta\gamma^\mu(1+\gamma_5)q_\beta],\;\; %\nonumber\\
Q_6= [\bar d_\alpha\gamma_\mu(1-\gamma_5)s_\beta]\sum_q[\bar q_\beta\gamma^\mu(1+\gamma_5)q_\alpha],\nonumber\\
&&Q_{12}=\frac{e}{16\pi^2}m_s\bar d\sigma^{\mu\nu}(1+\gamma_5)s F_{\mu\nu},\quad \tau=-\frac{V_{ts}V^*_{td}}{V_{us}V^*_{ud}},\label{H}
\end{eqnarray}
where the $z_i$ and $y_i$ are the corresponding Wilson coefficient. In our work the QCD peguin operators $Q_{3-6}$ are omitted, since its contribution is much smaller than current-current operator $Q_{1,2}$~\cite{Buchalla:1995vs}.

Recently, a new type of light-front quark model (LFQM) method based on the three-quark picture has been presented~\cite{Zhao:2023yuk}.  This model can express a baryon state in a more natural way, where the three quarks in the baryon are represented with three spinors $u$ and a charge conjugate transformation ${\mathcal C}$, which is different from other LFQM approaches in the diquark picture~\cite{Wang:2022ias,Liu:2022mxv}.  The model is easy to understand and has also been confirmed by their previous work on the $\Lambda_b\to\Lambda_c$ process.
 It's worth noting that the three-quark picture has also applied to the b-baryon decay cases~~\cite{Ke:2019smy,Li:2022nim,Li:2022hcn,Lu:2023rmq} because it provides a better description of the interaction between two constituent quarks in a baryon.  Additionally, it has a unique advantage in analyzing the baryon mixing problem and weak radiative decay processes~\cite{Deng:2023qaf}, where the QED effect is able to be included. 
 Nevertheless,  the applications and evaluations of three-quark picture have not been fully explored especially  for   light baryons in LFQM, which usually appear as the daughter states in heavy baryon decay processes. 
In our work, we will focus on weak radiative decay processes of light baryon by following the method outlined in Ref.~\cite{Zhao:2023yuk}.

With the help of this powerful method, we are able to estimate the contributions of the W exchange diagram shown in Fig.~\ref{fig1}. Based on the effective Hamiltonian mentioned above, we will investigate the following hyperon weak radiative decay processes:
\begin{eqnarray}
\Sigma^+\to p\gamma,\quad \Lambda\to n\gamma, \quad \Sigma^0\to n \gamma, \;\;\; \Xi^0\to\Lambda\gamma,\quad \Xi^0\to\Sigma^0\gamma, \quad \Xi^-\to \Sigma^- \gamma.
\end{eqnarray}
The remainder of this paper is organized as follows. We present the theoretical framework of LFQM in Section II.  Section III explicitly calculates the hyperon weak radiative decay processes   and provides the analytical expressions.  Section IV includes numerical analysis and discussions. Finally, we provide a brief summary.

%%%%%%%%%%%%%%%%%
\section{theoretical framework}
%%%%%%%%%%%%%%%%

When considering a baryon composed of three quarks, the three-quark picture LFQM can represent it using three-quark states as
\begin{eqnarray}
|{\mathcal B}(P,S,S_z)\rangle&=&\int\{d^3\tilde p_1\}\{d^3\tilde p_2\}\{d^3\tilde p_3\}
2(2\pi)^3\delta^3(\tilde P-\tilde p_1-\tilde p_2-\tilde p_3)\frac{1}{\sqrt{P^+}}\notag\\
%&\times&2(2\pi)^3\delta^3(\tilde P-\tilde p_1-\tilde p_2-\tilde p_3)\frac{1}{\sqrt{P^+}}\notag\\
&\times&\sum_{\lambda_1,\lambda_2,\lambda_3}\Psi^{S,S_z}(\tilde p_1,\tilde p_2,\tilde p_3,\lambda_1,\lambda_2,\lambda_3)C^{ijk}%\notag\\
|q^i_1(p_1,\lambda_1)q^j_2(p_2,\lambda_2)q^k_3(p_3,\lambda_3)\rangle.
%&\times&|q^i_1(p_1,\lambda_1)q^j_2(p_2,\lambda_2)q^k_3(p_3,\lambda_3)\rangle.
\label{baryons}
\end{eqnarray}
Here, $p_i(\lambda_i)$ represents the light-front momentum (helicity) of the quark denoted by $i$, and $C^{ijk}=\epsilon^{ijk}/\sqrt{6}$ represents the color wave function.  The wave function in both spin and momentum spaces is contained in $\Psi^{S,S_z}$. All the calculations in LFQM will be performed in the light-front frame, with the conventions:
\begin{eqnarray}
\tilde p_i=(p^+_i,p_{i\perp}),p^-=\frac{m^2_i+p_{i\perp}^2}{p_i^+}, \{d^3\tilde p_i\}=\frac{dp^+_id^2p_{i\perp}}{2(2\pi)^3}.
\end{eqnarray}
The four-momentum in the light-front frame is decomposed into $p_i=(p^-_i,p^+_i, p_{i\perp })$.
To describe the interaction between three quarks, we introduce intrinsic variables $(x_i,k_{i\perp })$ through
\begin{eqnarray}
&&p^+_i=x_iP^+,\quad p_{i \perp }=x_iP_\perp+k_{i\perp},\;\;\; %\notag\\
\sum_i x_i=1,\quad\sum_i k_{i\perp}=0,
\end{eqnarray}
Here, $x_i$ represents the light-front momentum fraction.  We can also define the invariant mass $M^2_0=\bar P^2$, where $\bar P=p_1+p_2+p_3$ and note that the momenta of each quark and baryon cannot be on their respective mass shells simultaneously. The mass $M_0$ can also be expressed in terms of $m_i, k_{i\perp}$, and $x_i$ in the rest frame of $\bar P$ as
\begin{eqnarray}
M_0^2=\sum_i\frac{k^2_{i\perp}+m_i^2}{x_i}.
\end{eqnarray}
And the momentum of quark can be expressed as
\begin{eqnarray}
&&p_i=(p_i^-,p_i^+,p_{i\perp})=(e_i-k_{iz},e_i+k_{iz},k_{i\perp})%\notag\\
=\left(\frac{m^2_i+k^2_{i\perp}}{x_iM_0},x_iM_0,k_{i\perp}\right),\notag\\
&&e_i=\frac{x_iM_0}{2}+\frac{m^2_i+k^2_{i\perp}}{2x_iM_0},k_{iz}=\frac{x_iM_0}{2}-\frac{m^2_i+k^2_{i\perp}}{2x_iM_0},
\end{eqnarray}
where $e_i$ represent the energy of the $i$-th quark.

 %%%%%%%%%%%%%%%%%%%%%%%%
\subsection{Spin and flavor wave function }
%%%%%%%%%%%%%%%%%%%%%%%%

\begin{figure}[htbp!]
	\includegraphics[width=0.4\columnwidth]{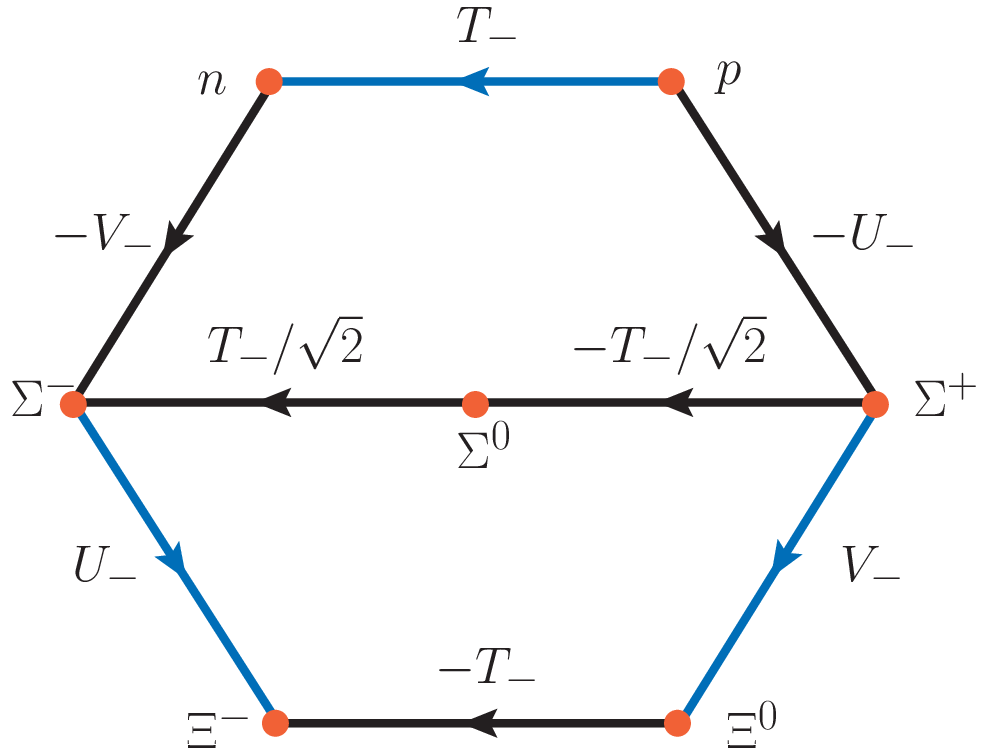}
	\caption{ The relation between light baryon octet under the iospin transformation $T_-,U_-,V_-$. }
\label{tran}
\end{figure}

Regarding the wave function $\Psi^{S,S_z}$, it is evidently more complex than the wave function of heavy baryons constructed in Ref.~\cite{Zhao:2023yuk}. To construct the wave function of a light baryon, we can begin by analyzing its spin wave function firstly.  Taking the $s_z=1/2$ baryon as an example, the spin wave functions are
\begin{eqnarray}
M_S&=&\frac{1}{\sqrt{6}}(2\uparrow\uparrow\downarrow-\uparrow\downarrow\uparrow-\downarrow\uparrow\uparrow),\;\;\;%\notag\\
M_A=\frac{1}{\sqrt{2}}(\uparrow\downarrow\uparrow-\downarrow\uparrow\uparrow).\label{wf}
\end{eqnarray}
Since the $s=1/2$ particle can be represented as a doublet representation of the SU(2) group, the total spin of a baryon can be seen as the representation from the direct product of three doublet representations  $2\otimes2\otimes2=2\oplus2\oplus4$, each corresponding to the spin of each quark.   This decomposition can be simplified in two steps. In the first step, the spins of two quarks are coupled into a singlet $s=0$ and a triplet $s=1$. Then, these two states are finally coupled to the last quark with spin-$\frac{1}{2}$.
This details for decomposition are shown as
\begin{eqnarray}
2\otimes2\otimes2&=&(2\otimes2)\otimes2=(1\oplus3)\otimes2%\notag\\
=(1\otimes2)\oplus(3\otimes2)=2\oplus2\oplus4.
\end{eqnarray}
Therefore, the symmetry wave function $M_S$ and anti-symmetry one $M_A$ in Eq.~\eqref{wf} can be considered as the doublet representation  respectively from the decomposition of the direct product of a triplet and a doublet, or  a singlet and a doublet.    As discussed in Ref.~\cite{Zhao:2023yuk}, the wave function $\Psi^{S,S_z}$, which corresponds to the symmetry and anti-symmetry spin wave functions, can be represented as
\begin{eqnarray}
M^s_S %:  \Psi_S^{S=\frac{1}{2},S_z}
&=&A_1\Phi(x_i,k_{i\perp}) \bar u(p_1,\lambda_1)\left(\frac{1}{\sqrt{3}}\gamma_\mu\gamma_5\right)u(\bar P,S_z)%\notag\\
\times \bar u(p_3,\lambda_3)(\bar{\slashed P}+M_0)(\gamma^\mu-v^\mu)C\bar u^T(p_2,\lambda_2),\notag\\
M^s_A %: \Psi_A^{S=\frac{1}{2},S_z}
&=& A_1\Phi(x_i,k_{i\perp}) \bar u(p_1,\lambda_1)u(\bar P,S_z)%\notag\\
\times\bar u(p_3,\lambda_3)(\bar{\slashed P}+M_0)(-\gamma_5)C\bar u^T(p_2,\lambda_2),\label{HFWF}
\end{eqnarray}
where $v^\mu=\bar P^\mu/M_0$.  These wave functions are sufficient to describe the spin and flavor of heavy baryons, as these twos  are decoupled from the heavy quark due to heavy quark symmetry.  However, in our work for light baryons, flavor symmetry should further be considered.  Taking the proton of the octuplet light baryon as an example, the symmetry and anti-symmetry flavor wave functions are
\begin{eqnarray}
M^f_S&:&  \frac{1}{\sqrt{6}}(2uud-udu-duu),\;\;\; %\notag\\
M^f_A:  \frac{1}{\sqrt{2}}(udu-duu)\label{fwf1}.
\end{eqnarray}
Combining Eq.\eqref{HFWF} and Eq.\eqref{fwf1}, the wave function of a proton can be written as $\frac{1}{\sqrt{2}}(M^s_S\times M^f_S+M^s_A\times M^f_A)$.  Following this method, we can  construct our wave function $\Psi^{S,S_z}$.  In the first step, we can express the wave function $\Psi_{S/A}^{S,S_z}$ in Eq.~\eqref{HFWF} as
\begin{eqnarray}
% \Psi_{S/A}^{S,S_z}&=&
 \bar u(p_3,\lambda_3)_\alpha \bar u^T(p_2,\lambda_2)_\beta\bar u(p_1,\lambda_1)_\gamma 
 %\notag\\
% &&
  \Gamma_{S/A}^{\alpha\beta\gamma}A_1\Phi(x_i,k_{i\perp}),
\end{eqnarray}
with
\begin{eqnarray}
 \Gamma_{S}^{\alpha\beta\gamma} &=&[(\bar{\slashed P}+M_0)(\gamma^\mu-v^\mu)C]^{\alpha\beta} \left[\frac{1}{\sqrt{3}}\gamma_\mu\gamma_5 u(\bar P,S_z)\right]^\gamma,\notag\\
  \Gamma_{A}^{\alpha\beta\gamma} &=&[(\bar{\slashed P}+M_0)(-\gamma_5)C]^{\alpha\beta} [u(\bar P,S_z)]^\gamma,
 \end{eqnarray}
  If we consider the flavor symmetry, the wave function $\Psi^{S,S_z}$ of light baryon can be constructed as
 \begin{eqnarray}
 \Psi^{S,S_z}&=&\frac{1}{\sqrt{2}}A_1\Phi(x_i,k_{i\perp})(M^S_{\alpha\beta\gamma}\Gamma_{S}^{\alpha\beta\gamma}+M^A_{\alpha\beta\gamma}\Gamma_{A}^{\alpha\beta\gamma}),\notag\\
 M^S_{\alpha\beta\gamma}&=&\frac{1}{\sqrt{6}}\bigg(2\bar u(p_3,\lambda_3)_\alpha \bar u(p_2,\lambda_2)_\beta\bar d(p_1,\lambda_1)_\gamma%\notag\\
 -\bar u(p_3,\lambda_3)_\alpha \bar d(p_2,\lambda_2)_\beta\bar u(p_1,\lambda_1)_\gamma%\notag\\
-\bar d(p_3,\lambda_3)_\alpha \bar u(p_2,\lambda_2)_\beta\bar u(p_1,\lambda_1)_\gamma\bigg),\notag\\
 M^A_{\alpha\beta\gamma}&=&\frac{1}{\sqrt{2}}\bigg(\bar u(p_3,\lambda_3)_\alpha \bar d(p_2,\lambda_2)_\beta\bar u(p_1,\lambda_1)_\gamma%\notag\\
 %&& \qquad
 -\bar d(p_3,\lambda_3)_\alpha \bar u(p_2,\lambda_2)_\beta\bar u(p_1,\lambda_1)_\gamma\bigg),
 \end{eqnarray}
Here, we use $u(p,\lambda)$ and $d(p,\lambda)$ to represent the spinors for different flavors. 
It seems we have obtained the wave function that satisfies both the spin and flavor symmetries.
%However, this form is still not complete because flavor symmetry should also apply to the state 
 %$|q^i_1(p_1,\lambda_1)q^i_2(p_2,\lambda_2)q^i_3(p_3,\lambda_3)\rangle$ in Eq.~\eqref{baryons}.
 To simplify these wave functions and the states in LFQM, we can define $p_i$ as the momentum of a quark with a specified flavor, such as
 \begin{eqnarray}
|{\mathcal B}(P,S,S_z)\rangle&=&\int\{d^3\tilde p_1\}\{d^3\tilde p_2\}\{d^3\tilde p_3\}%\frac{1}{\sqrt{P^+}}
%\notag\\
\times 2(2\pi)^3\delta^3(\tilde P-\tilde p_1-\tilde p_2-\tilde p_3)\frac{1}{\sqrt{P^+}}\notag\\
&\times&\sum_{\lambda_1,\lambda_2,\lambda_3}\Psi^{S,S_z}(\tilde p_1,\tilde p_2,\tilde p_3,\lambda_1,\lambda_2,\lambda_3)C^{ijk}%\notag\\
 |u^i(p_3,\lambda_1)u^j(p_2,\lambda_2)d^k(p_1,\lambda_3)\rangle.
\end{eqnarray}
Since the $\Phi(x_i,k_{i\perp})$ in wave function is symmetry with  $x_i$ exchange, we can write the wave function $\Psi^{SS_z}$ as
\begin{eqnarray}
 \Psi^{S,S_z}&=&
 \frac{1}{\sqrt{2}}A_1\Phi(x_i,k_{i\perp})\bar u(p_3,\lambda_3)_\alpha \bar u(p_2,\lambda_2)_\beta\bar u(p_1,\lambda_1)_\gamma %\notag\\
 \times\bigg(\frac{1}{\sqrt{6}}(2\Gamma_{S}^{\alpha\beta\gamma}-\Gamma_{S}^{\alpha\gamma\beta}-\Gamma_{S}^{\beta\gamma\alpha})
% \notag\\
% &&
+\frac{1}{\sqrt{2}}(\Gamma_{A}^{\alpha\gamma\beta}-\Gamma_{A}^{\gamma\beta\alpha})\bigg).\notag\\\label{fwf}
 \end{eqnarray}
A fierz transformation of dirac structure $\Gamma_{S/A}^{\alpha\beta\gamma}$ can be applied to simplify the wave function. The results by  the fierz transformation are
\begin{eqnarray}
\Gamma_{S}^{\alpha\gamma\beta}&=&\frac{1}{4\sqrt{3}}\bigg(\frac{1}{4}[(\bar{\slashed P}+M_0)\sigma^{\mu\nu}\gamma_5C]^{\alpha\beta}[\sigma_{\mu\nu}u(\bar P,S_z)]^\gamma%\notag\\
%&&\quad \quad \quad
-[(\bar{\slashed P}+M_0)(\gamma^\mu-v^\mu)C]^{\alpha\beta}[\gamma_\mu\gamma_5u(\bar P,S_z)]^\gamma\notag\\
&&\quad \quad \quad
+6[(\bar{\slashed P}+M_0)\gamma_5C]^{\alpha\beta}[u(\bar P,S_z)]^\gamma\bigg)\notag\\
\Gamma_{S}^{\beta\gamma\alpha}&=&\frac{1}{4\sqrt{3}}\bigg(\frac{1}{4}[(\bar{\slashed P}+M_0)\sigma^{\mu\nu}\gamma_5C]^{\alpha\beta}[\sigma_{\mu\nu}u(\bar P,S_z)]^\gamma%\notag\\
%&&\quad \quad \quad
-[(\bar{\slashed P}+M_0)(\gamma^\mu-v^\mu) C]^{\alpha\beta}[\gamma_\mu \gamma_5 u(\bar P,S_z)]^\gamma\notag\\
&&\quad \quad \quad-[M_0\gamma^\mu\gamma_5 C]^{\alpha\beta}[\gamma_\mu u(\bar P,S_z)]^\gamma%\notag\\
%&&\quad \quad \quad
-[(5\bar{\slashed P}+6M_0)\gamma_5C]^{\alpha\beta}[u(\bar P,S_z)]^\gamma\bigg)\label{fies}
 \end{eqnarray}
and
\begin{eqnarray}
\Gamma_{A}^{\alpha\gamma\beta}&=&\frac{-1}{4}\bigg(\frac{-1}{4}[(\bar{\slashed P}+M_0)\sigma^{\mu\nu}\gamma_5C]^{\alpha\beta}[\sigma_{\mu\nu}u(\bar P,S_z)]^\gamma%\notag\\
%&&\quad \quad \quad
+[(\bar{\slashed P}+M_0)(\gamma^\mu+v^\mu)\gamma_5C]^{\alpha\beta}[\gamma_\mu u(\bar P,S_z)]^\gamma\notag\\
&&\quad \quad \quad+[(\bar{\slashed P}+M_0)(\gamma^\mu-v^\mu)C]^{\alpha\beta}[\gamma_\mu\gamma_5u(\bar P,S_z)]^\gamma \bigg)\notag\\
\Gamma_{A}^{\beta\gamma\alpha}&=&\frac{1}{4}\bigg(\frac{1}{4}[(\bar{\slashed P}+M_0)\sigma^{\mu\nu}\gamma_5C]^{\alpha\beta}[\sigma_{\mu\nu}u(\bar P,S_z)]^\gamma%\notag\\
%&&\quad \quad 
-[(\bar{\slashed P}+M_0)(\gamma^\mu-v^\mu) C]^{\alpha\beta}[\gamma_\mu \gamma_5 u(\bar P,S_z)]^\gamma\notag\\
&&\quad \quad-[\bar{\slashed P}(\gamma^\mu+v^\mu)\gamma_5 C]^{\alpha\beta}[\gamma_\mu u(\bar P,S_z)]^\gamma%\notag\\
%&&\quad \quad
+[3(\bar{\slashed P}+M_0)\gamma_5C]^{\alpha\beta}[u(\bar P,S_z)]^\gamma\bigg)\label{fiea}
 \end{eqnarray}
 Combining Eq.~\eqref{fwf}, Eq.~\eqref{fies} and Eq.~\eqref{fiea}, we can derive the completely wave function $\Psi^{SS_z}$ of proton as
 \begin{eqnarray}
 \Psi^{S,S_z}_p&=&
\bar u(p_3,\lambda_3)_\alpha \bar u(p_2,\lambda_2)_\beta\bar u(p_1,\lambda_1)_\gamma %\notag\\
 \Gamma^{\alpha\beta\gamma}_p A_p\Phi(x_i,k_{i\perp})\label{cwf}
    \end{eqnarray}
with 
  \begin{eqnarray}
\Gamma^{\alpha\beta\gamma}_p&=&\bigg([(\bar{\slashed P}+M_0)\sigma^{\mu\nu}\gamma_5C]^{\alpha\beta}[\sigma_{\mu\nu} u(\bar P,S_z)]^\gamma%\notag\\
% &&\quad 
 +4[(\bar{\slashed P}+M_0)(\gamma^\mu-v^\mu) C]^{\alpha\beta}[\gamma_\mu\gamma_5 u(\bar P,S_z)]^\gamma\notag\\
 && \quad -2[(3\bar{\slashed P}+M_0)\gamma^\mu\gamma_5C]^{\alpha\beta}[\gamma_\mu u(\bar P,S_z)]^\gamma%\notag\\
% && \quad  
 +2[(\bar{\slashed P}+3M_0)\gamma_5C]^{\alpha\beta}[ u(\bar P,S_z)]^\gamma\bigg).\label{wave}
 \end{eqnarray}
Here  $A_p$ is the normalization coefficient of proton. % with $A_p=A_1/24$.
As for the other light baryon octets, they can be obtained by  using  the lowering operators on  proton state. Following the method in Ref.~\cite{Bali:2015ykx},   the lowering operators $T_-,U_-,V_-$ are defined as 
 \begin{eqnarray}
 T_-u=d,\quad U_-d=s,\quad V_-u=s.
  \end{eqnarray}
Then, starting from the proton state, the
light baryon octet can be constructed by applying the following transformations as shown  in Fig.~\ref{tran}:
 \begin{eqnarray}
&&T_-|p\rangle=|n\rangle,\; -U_-|p\rangle=|\Sigma^+\rangle,\; \frac{1}{\sqrt{2}}T_-U_-|p\rangle=|\Sigma^0\rangle
\frac{1}{2}T_-T_-U_-|p\rangle=|\Sigma^-\rangle,\;\notag\\
&&-V_-U_-|p\rangle=|\Xi^0\rangle,%\notag\\
T_-V_-U_-|p\rangle=|\Xi^-\rangle,\; \frac{-1}{\sqrt{6}}(V_-+U_-T_-)|p\rangle=|\Lambda\rangle.\label{iospin}
  \end{eqnarray}

  Following the method in Ref.~\cite{Wein:2015oqa,Bali:2015ykx} and applying the Eq.~\eqref{iospin} in proton wave function, one can derive the expression of baryon state similar to  Eq.~\eqref{baryons} with  each baryon flavor  $\{q_3,q_2,q_1\}$  as
   \begin{eqnarray}
p:\;uud,\quad n:\;ddu,\quad \Sigma^+:\;uus,\quad\Sigma^0:\;uds,%\notag\\
\Sigma^-:\;dds,\quad\Xi^0:\;ssu,\quad \Xi^-:\;ssd,\quad \Lambda:\; uds.\label{fo}
  \end{eqnarray}  
  And  $\Gamma^{\alpha\beta\gamma}_B$ in wave function $ \Psi^{S,S_z}$ have the relations below
   \begin{eqnarray}
\Gamma_N^{\alpha\beta\gamma}&\equiv &\Gamma_p^{\alpha\beta\gamma}=- \Gamma_n^{\alpha\beta\gamma},\;\;%\notag\\
\Gamma^{\alpha\beta\gamma}_\Sigma\equiv\Gamma^{\alpha\beta\gamma}_{\Sigma^-}=\sqrt{2} \Gamma^{\alpha\beta\gamma}_{\Sigma^0}=-\Gamma^{\alpha\beta\gamma}_{\Sigma^+},\;\;%\notag\\
\Gamma^{\alpha\beta\gamma}_\Xi\equiv\Gamma^{\alpha\beta\gamma}_{\Xi^0}=-\Gamma^{\alpha\beta\gamma}_{\Xi^-},
  \end{eqnarray}
  where we divide the light baryon octet into four part: $(N,\Sigma,\Xi,\Lambda)$.  By analyzing the flavor symmetry of baryon in $(N,\Sigma,\Xi)$ type, one can also find 
     \begin{eqnarray}
\Gamma_N^{\alpha\beta\gamma}=\Gamma^{\alpha\beta\gamma}_\Sigma=\Gamma^{\alpha\beta\gamma}_\Xi.
  \end{eqnarray}
Similarly,  the  $\Lambda$ wave function  $\Psi^{S,S_z}_\Lambda$ can be obtained as 
 \begin{eqnarray}
	\bar u(p_3,\lambda_3)_\alpha \bar u(p_2,\lambda_2)_\beta\bar u(p_1,\lambda_1)_\gamma 
	\Gamma^{\alpha\beta\gamma}_\Lambda A_\Lambda\Phi(x_i,k_{i\perp}),
 \end{eqnarray}
 with
  \begin{eqnarray}
	\Gamma^{\alpha\beta\gamma}_\Lambda&=&[M_0(\gamma^\mu+v^\mu)\gamma_5C]^{\alpha\beta}[\gamma_\mu u(\bar P,S_z)]^\gamma%\notag\\
	%&& \quad 
	 +11[(\bar{\slashed P}+M_0)\gamma_5C]^{\alpha\beta}[ u(\bar P,S_z)]^\gamma.\label{lambda}
\end{eqnarray}
Here  $A_\Lambda$ is the normalization coefficient of proton. %with $A_\Lambda=A_1/4\sqrt{6}$.
  
   %%%%%%%%%%%%%%%%%%%%%%%%
\subsection{Internal momenta distribution function }
%%%%%%%%%%%%%%%%%%%%%%%%

In addition to  the spin and flavor wave function, the distribution of internal momenta $k_i$ are also described by the wave function $\Phi(x_i,k_{i\perp})$ and  usually  expressed as
\begin{eqnarray}
\Phi(x_i,k_{i\perp})&=&\sqrt{\frac{e_1e_2e_3}{x_1x_2x_3M_0}}16\left(\frac{\pi}{\beta^2}\right)^{3/2}\pi^{-3/16}%\notag\\
\times\exp\left[-\sum^3_{i\leq j}\frac{\vec{k}_{i\perp}\cdot \vec{k}_{j\perp}+k_{iz}k_{jz}}{2\beta^2}\right],\label{md}
\end{eqnarray}
%&\times&\exp{\frac{-(\vec{k}_1\vec{k}_2+\vec{k}_3\vec{k}_2+\vec{k}_1\vec{k}_3)-(k_{z1}k_{z2}+k_{z3}k_{z2}+k_{z1}k_{z3})}{2\beta^2}}\notag\\
%&\times&\exp\left[\frac{-(\vec{k}_1^2+\vec{k}_2^2+\vec{k}_3^2)-(k_{z1}^2+k_{z2}^2+k_{z3}^2)}{2\beta^2}\right],\label{md}
%&\times&\exp\left[-\sum^3_{i=1}\frac{\vec{k}_{i\perp}^2+k_{iz}^2}{2\beta^2}\right],\label{md}
%\end{eqnarray}
where  $\beta$ is the shape parameter. 
The coefficient $A_1$ in Eq.~\eqref{cwf} is normalized parameter  determined by normalizing the baryon state
     \begin{eqnarray}
&&\langle B(P^\prime,S^\prime,S^\prime_z)|B(P,S,S_z)\rangle%\notag\\
=2(2\pi)^3P^+\delta_{S^\prime S}\delta_{S_z^\prime S_z}
 \delta^3(\tilde{P}^\prime-\tilde{P}),
  \end{eqnarray}
  with
       \begin{eqnarray}
&&\int \Bigg( \prod_{i=1}^3 \frac{dx_id^2k_{i\perp}}{2(2\pi)^3}\Bigg)2(2\pi)^3 |\Phi(x_i,k_{i\perp})|^2%\notag\\
%&&\qquad \qquad 
\times \delta\left(1-\sum x_i\right)\delta^2\left(\sum k_{i\perp}\right) =1.
  \end{eqnarray}
One can derive that 
  \begin{eqnarray}
    A_B&=&\frac{1}{2} A_{\Sigma^0}%\notag\\
    =\frac{1}{8\sqrt{6 M_0^3(e_1+m_1)(e_2+m_2)(e_3+m_3)}},\;\;\;%\notag\\
    A_{\Lambda}=\frac{1}{48\sqrt{ M_0^3(e_1+m_1)(e_2+m_2)(e_3+m_3)}}.
  \end{eqnarray}
According to the discussion in Ref.~\cite{Zhao:2023yuk}, the shape parameter is defined as the normalization constant in the baryon light-cone distribution amplitude,  which can be determined by the baryon state decay constant~\cite{RQCD:2019hps} from the local matrix element with the definitions as 
\begin{eqnarray}
&&\langle 0|[f(0)^TC\gamma^\mu(1-\gamma_5)g(0)]\gamma_\mu(1+\gamma_5) h(0)|(B\neq\Lambda)_{p,\lambda}\rangle%\notag\\
%&&\qquad\qquad
=-\lambda^B_1 m_B (1-\gamma_5)u^B(p,\lambda),\notag\\
&&\langle 0|[u(0)^TC\gamma^\mu(1-\gamma_5) d(0)]\gamma_\mu (1+\gamma_5)s(0)|\Lambda_{p,\lambda}\rangle%\notag\\
%&&\qquad\qquad
=\frac{1}{\sqrt{6}}\lambda_1^\Lambda m_\Lambda (1-\gamma_5)u^\Lambda(p,\lambda),\label{dc}
\end{eqnarray}
where the flavor order in each baryon octet obeys Eq.\eqref{fo}. The decay constants of the baryon state  $\lambda_1^B$ are evaluated by using Lattice QCD~\cite{RQCD:2019hps} as shown in Table.~\ref{table1}.

Applying Eq.~\eqref{baryons} into the definition of decay constant, we can obtain
\begin{eqnarray}
&&K_{B\neq\Lambda}^{\alpha\beta\gamma}[C(\slashed p_3-m_3)\gamma^\mu(1-\gamma_5)(\slashed p_2+m_2)]_{\alpha\beta}%\notag\\
%&&\qquad \quad  
\times [\gamma_\mu(1+\gamma_5)(\slashed p_1+m_1)]_\gamma
%\notag\\
%&&\qquad 
=-\lambda^B_1m_B(1-\gamma_5)u^{B\neq\Lambda}(p,\lambda),\notag\\
&&K_\Lambda^{\alpha\beta\gamma}[C(\slashed p_3-m_u)\gamma^\mu(1-\gamma_5)(\slashed p_2+m_d)]_{\alpha\beta}%\notag\\
%&&\qquad \quad
\times[\gamma_\mu(1+\gamma_5)(\slashed p_1+m_s)]_\gamma%\notag\\
%&&\qquad 
=\frac{1}{\sqrt{6}}\lambda^\Lambda_1m_\Lambda(1-\gamma_5)u^{\Lambda}(p,\lambda).\\ \label{beta}
&& {\rm with}\quad K^{\alpha\beta\gamma}=\int\{d^3\tilde p_1\}\{d^3\tilde p_2\}\{d^3\tilde p_3\}%\notag\\
% &&\qquad\qquad
 \times2(2\pi)^3\delta^3(\tilde P-\sum_i\tilde p_i)\frac{A_B \Phi(x_i,k_{i\perp})\Gamma^{\alpha\beta\gamma}}{(P^+)^2\sqrt{x_1x_2x_3}}\notag.
\end{eqnarray}
After multiplying the structure $\sum_\lambda \bar u(p,\lambda)$ in both two sides of Eq.~\eqref{beta}, the decay constant can be expressed in our model as
\begin{widetext}
\begin{eqnarray}
\lambda^{B\neq\Lambda}_1&=&\frac{1}{4 m_B^2}\int \frac{dx_2d^2\vec k_{2\perp}}{2(2\pi)^3}\frac{dx_3d^2\vec k_{3\perp}}{2(2\pi)^3}\frac{1}{\sqrt{x_1x_2x_3}}A_{B\neq \Lambda} \Phi(x_i,k_{i\perp})\notag\\
&\times& \bigg(2{\rm Tr}[(\bar {\slashed P}+3M_0)\gamma_5(\slashed p_2-m_2)(1-\gamma_5)\gamma^\mu(\slashed p_3+m_3)] {\rm Tr}[\gamma_\mu(1+\gamma_5)(\slashed p_1+m_1)(\bar {\slashed P}+M_0)]\notag\\
&&\quad-2{\rm Tr}[(3\bar {\slashed P}+M_0)\gamma^\alpha\gamma_5(\slashed p_2-m_2)(1-\gamma_5)\gamma^\mu(\slashed p_3+m_3)] {\rm Tr}[\gamma_\mu(1+\gamma_5)(\slashed p_1+m_1)\gamma_\alpha(\bar {\slashed P}+M_0)]\notag\\
&&\quad+4{\rm Tr}[(\bar {\slashed P}+M_0)(\gamma^\alpha-v^\alpha)(\slashed p_2-m_2)(1-\gamma_5)\gamma^\mu(\slashed p_3+m_3)] {\rm Tr}[\gamma_\mu(1+\gamma_5)(\slashed p_1+m_1)\gamma_\alpha\gamma_5(\bar {\slashed P}+M_0)]\notag\\
&&\quad+{\rm Tr}[(\bar {\slashed P}+M_0)\sigma^{\alpha\beta}\gamma_5(\slashed p_2-m_2)(1-\gamma_5)\gamma^\mu(\slashed p_3+m_3)] {\rm Tr}[\gamma_\mu(1+\gamma_5)(\slashed p_1+m_1)\sigma_{\alpha\beta}(\bar {\slashed P}+M_0)]\bigg),\notag\\
\lambda^\Lambda_1&=&-\frac{1}{4 m^2_\Lambda}\int \frac{dx_2d^2\vec k_{2\perp}}{2(2\pi)^3}\frac{dx_3d^2\vec k_{3\perp}}{2(2\pi)^3}\frac{\sqrt{6}}{\sqrt{x_1x_2x_3}}A_\Lambda \Phi(x_i,k_{i\perp})\notag\\
&\times&\bigg(11{\rm Tr}[(\bar {\slashed P}+M_0)\gamma_5(\slashed p_2-m_d)(1-\gamma_5)\gamma^\mu(\slashed p_3+m_u)] {\rm Tr}[\gamma_\mu(1+\gamma_5)(\slashed p_1+m_s)(\bar {\slashed P}+M_0)]\notag\\
&&+{\rm Tr}[M_0(\gamma^\alpha+v^\alpha)\gamma_5(\slashed p_2-m_d)(1-\gamma_5)\gamma^\mu(\slashed p_3+m_u)] {\rm Tr}[\gamma_\mu(1+\gamma_5)(\slashed p_1+m_s)\gamma_\alpha(\bar {\slashed P}+M_0)]\bigg).
\end{eqnarray}
\end{widetext}
Based on these formulas, we can estimate the value of the shape parameter $\beta$ for each light baryon as  shown in Table.~\ref{table1}.

%%%%%%%%%%%%%%%%%%%%%%%%%%%%%%%%%%%%
%\begin{widetext}
\begin{center}
 \begin{table}[!htb]
\caption{The value of decay constant $\lambda^B_1$ of baryon octet defined in Eq.~\eqref{dc} in LQCD and the determined value of shape parameter $\beta_1/\beta_{23}$ in Eq.~\eqref{md}.}
\label{table1}%
\begin{tabular}{cccccccccc}
\hline \hline
 baryons  & \;\;$N$\qquad \qquad &\; \;$\Sigma$ \qquad\qquad&\;\;$\Xi$\qquad\qquad&\;\;$\Lambda$\qquad \qquad\tabularnewline
\hline
$\lambda^B_1(10^{-3}{\rm GeV}^2)$ &-44.9&-46.1&-49.8&-42.2\tabularnewline
\hline
$\beta({\rm GeV})$&0.366&0.410&0.419& 0.378   \tabularnewline
\hline
\hline
\end{tabular}
\end{table}
\end{center}
%\end{widetext}

 %%%%%%%%%%%%%%%%%%%%%%%%
\section{Application of light-front quark model in three quark picture }
%%%%%%%%%%%%%%%%%%%%%%%%

\begin{widetext}
\begin{figure*}[htbp!]
	\begin{minipage}[t]{0.28\linewidth}
		\centering
		\includegraphics[width=1\columnwidth]{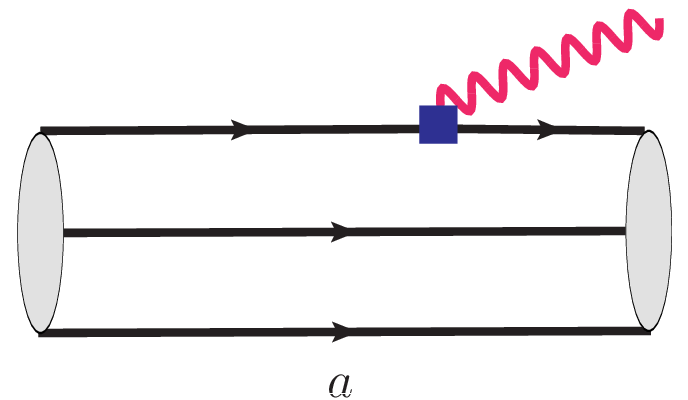}
	\end{minipage}
	\begin{minipage}[t]{0.28\linewidth}
		\centering
		\includegraphics[width=1\columnwidth]{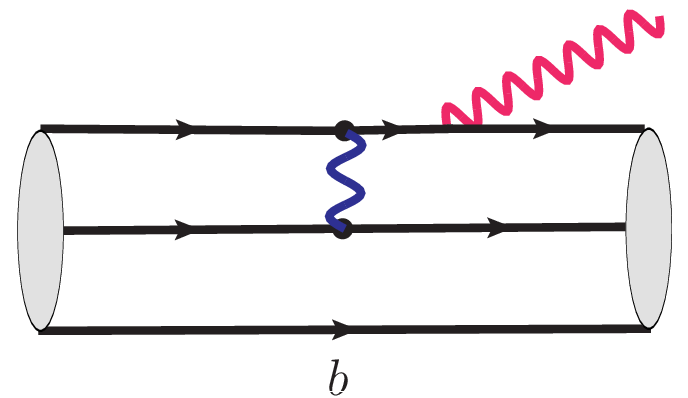}
	\end{minipage}
	\begin{minipage}[t]{0.28\linewidth}
		\centering
		\includegraphics[width=1\columnwidth]{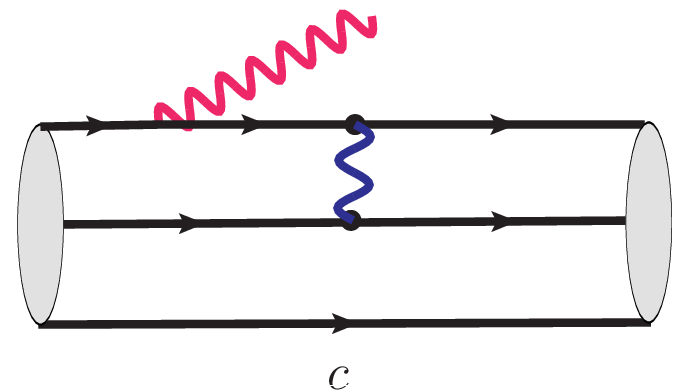}
	\end{minipage}
	\begin{minipage}[t]{0.28\linewidth}
		\centering
		\includegraphics[width=1\columnwidth]{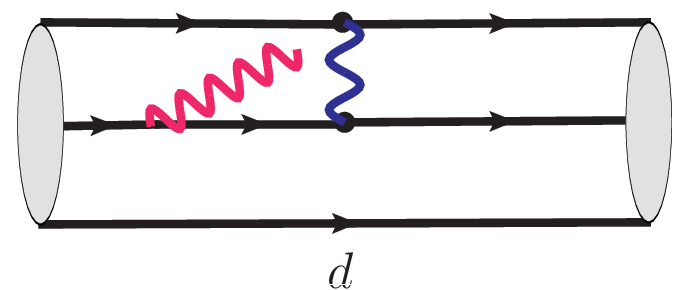}
	\end{minipage}
 \begin{minipage}[t]{0.28\linewidth}
		\centering
		\includegraphics[width=1\columnwidth]{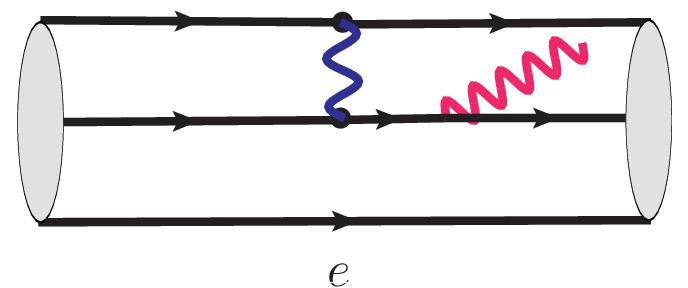}
	\end{minipage}
 \begin{minipage}[t]{0.28\linewidth}
		\centering
		\includegraphics[width=1\columnwidth]{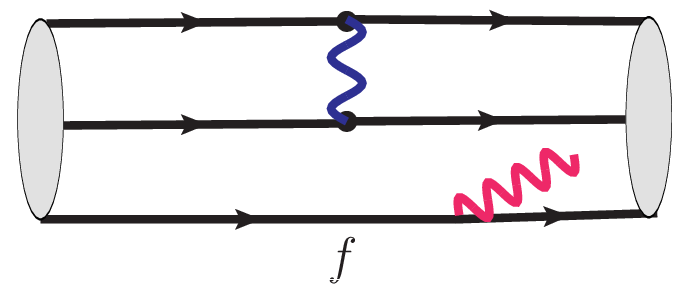}
	\end{minipage}
	\caption{The Feynman diagrams of radiative decay processes induced by the effective Hamiltonian in Eq.~\eqref{H}. Red (blue)  wavy lines means the photon (W boson). Fig(a) and Fig(b-f) correspond to the $Q_{12}$ and $Q_{1/2}$ operators, respectively.}
\label{fig1}
\end{figure*} 
\end{widetext}

After deriving the light-front quark model for the light baryon octet in the three-quark picture, we can calculate the radiative decay processes induced by the effective Hamiltonian in Eq.~\eqref{H}.
Since baryon states are described by three constituent quark states, the processes involving the W-boson exchange can be easily estimated. This is also one of the advantages of our model.  In the first step, the Feynman diagrams for hyperon weak radiative decays are depicted in Fig.~\ref{fig1}. 

%%%%%%%%%%%%%%%%%%%%%%%%%%%%%%%
%%%%%%%%%%%%%%%%%%%%%%%%%%%%%%%%%%%%%%%%%

There are two types of interactions that can contribute to these processes: penguin diagrams~Fig.~\ref{fig1}(a) and W exchange diagrams~Fig.~\ref{fig1}(b-f).  As we mentioned in Ref.~\cite{Shi:2022zzh}, the amplitudes of spin-$\frac{1}{2}$ hyperon radiative decays are described by a parity-conserving (P-wave) and a parity-violating (S-wave) amplitude.

 %%%%%%%%%%%%%%%%%%%%%%%%
\subsection{Form factors }
%%%%%%%%%%%%%%%%%%%%%%%%
The amplitude of the hyperon radiative decays can be expressed by two form factors as
\begin{eqnarray}
\langle  B^\prime\gamma|\mathcal{H}(0)|B\rangle=\frac{i G_F}{\sqrt{2}} \bar u_{B^\prime}(a+b\gamma_5)\sigma_{\mu\nu}u_B\frac{q^\nu \epsilon^{*\mu}}{m_{B}},
\end{eqnarray}
where $q=P-P^\prime$ is the momentum of radiated photon. And the coefficient a(b) means S-wave(p-wave) form factor.
It is evident that these two form factors can be extracted in terms of the baryon state within LFQM.
The matrix element can be obtained by the amplitude associated with six Feynman diagrams as
\begin{eqnarray}
 {\mathcal M}&\equiv& \langle  B^\prime\gamma|\mathcal{H}(0)|B\rangle
= \frac{G_F}{\sqrt{2}}({\mathcal M}_P+ {\mathcal M}_T),\label{am}
\end{eqnarray}
where the subscripts $P,T$ correspond to the penguin Feynman diagram in Fig.~\ref{fig1}$(a)$ and tree Feynman diagram in Fig.~\ref{fig1}$(b\sim f)$, respectively, and
\begin{eqnarray}
{\mathcal M}_T
&=& V_{us}V^*_{ud}\sum_i^{1,2}(z_{i}+\tau y_i)\langle B^\prime \gamma|Q_i|B\rangle,\;\;\;%\notag\\
{\mathcal M}_P=V_{us}V^*_{ud}(z_{12}+\tau y_{12})\frac{e}{8\pi^2}m_s%\notag\\
%&&
\times\langle B^\prime \gamma|\bar d\sigma^{\mu\nu}(1+\gamma_5)sF_{\mu\nu}|B\rangle,
\end{eqnarray}
In order to extract out the form factors, we follow the method in Ref.~\cite{Liu:2023zvh} to multiply the structure $\bar u_B\slashed \epsilon u_{B^\prime}$ (or $\bar u_B\slashed \epsilon\gamma_5 u_{B^\prime}$) and sum over their spins.  
Correspondingly,  the two amplitudes in LFQM are defined as $H=\sum_{s}\bar u_B\slashed \epsilon u_{B^\prime}\times{\mathcal M}$ and $H_5=\sum_s\bar u_B \slashed \epsilon\gamma_5u_{B^\prime}\times {\mathcal M}$.  Then the two form factors ($a$ and $b$) can be expressed in terms of $H$ and $H_5$ as
\begin{eqnarray}
a=-\frac{H M}{6(M+M^\prime)Q_-},\;b=\frac{H_5 M}{6(M-M^\prime)Q_+}\label{ff}.
\end{eqnarray}
where $Q_{\pm}=(M^\prime\pm M)^2-q^2$.  
To calculate the amplitudes $H_{(5)}$, we should consider the contribution from each Feynman diagram simultansously.
The amplitude ${\mathcal M}_{P}$  can be expressed as	
\begin{eqnarray}
{\mathcal M}_P&=&-V_{us}V^*_{ud}(z_{12}+\tau y_{12})\frac{e}{8\pi^2}m_s \Gamma_B^{abc}\overline \Gamma_{B^\prime}^{\alpha\beta\gamma}%\notag\\
%&\times &
\int \frac{dx_1d^2\vec k_{1\perp}}{2(2\pi)^3}\frac{dx_2d^2\vec k_{2\perp}}{2(2\pi)^3} \frac{\Phi(x_i,k_{i\perp})\Phi^*(x^\prime_i,k^\prime_{i\perp})}
{\sqrt{x_1(1-x^\prime_2-x^\prime_3)}}\notag\\
&\times&[(\slashed p'_1+m'_1)\sigma^{\mu\nu}(1+\gamma_5)(\slashed p_1+m_1)]_{\gamma c}%\notag\\
\times [\slashed p_3+m_3]_{\alpha a}[\slashed p_2+m_2]_{\beta b}A_B A_{B'} (2iq_\nu \epsilon^*_\mu(q)),
\end{eqnarray}
where  $p^{(\prime)}_1=\bar P^{(\prime)}-p_2-p_3$ and $q$ is the photon momentum. By taking the hermitian conjugate of the wave function $\Psi^{S,S_z}$,   $\overline \Gamma$ can be obtained as
 \begin{eqnarray}
\overline \Gamma^{\alpha\beta\gamma}_{B\neq\Lambda}&=&\bigg(-[C\gamma_5\sigma^{\mu\nu}(\bar{\slashed P}+M_0)]^{\beta\alpha}[\bar u(\bar P,S_z) \sigma_{\mu\nu} ]^\gamma%\notag\\
% &&\quad 
 +4[C (\gamma^\mu-v^\mu) (\bar{\slashed P}+M_0) ]^{\beta\alpha}[\bar u(\bar P,S_z)\gamma_\mu \gamma_5 ]^\gamma\notag\\
 && \quad +2[C\gamma_5 \gamma^\mu(3\bar{\slashed P}+M_0)]^{\beta\alpha}[\bar u(\bar P,S_z)\gamma_\mu ]^\gamma%\notag\\
 -2[C\gamma_5(\bar{\slashed P}+3M_0)]^{\beta\alpha}[ \bar u(\bar P,S_z)]^\gamma\bigg),\notag\\
 \overline \Gamma^{\alpha\beta\gamma}_{\Lambda}&=&\bigg(-[ M_0 C\gamma_5 (\gamma^\mu+v^\mu)]^{ \beta\alpha} [ \bar u(\bar P,S_z)\gamma_\mu]^\gamma %\notag\\
 -11[C\gamma_5 (\bar{\slashed P}+M_0) ]^{\beta\alpha}[ \bar u(\bar P,S_z)]^\gamma\bigg).\label{cwfbar}
 \end{eqnarray}

For the amplitude ${\mathcal M^i_{T}}(i=\{b\sim f\})$ , it can be expressed by LFQM as
\begin{eqnarray}
{\mathcal M}^b_T&=&C\int dx_1 dx_2dx^\prime_1 \frac{d^2k_{1\perp}}{2(2\pi)^3}\frac{d^2k_{2\perp}}{2(2\pi)^3}\frac{d^2k^\prime_{1\perp}}{2(2\pi)^3}%\notag\\
 \frac{A_B A_{B'} \epsilon^*_\mu(q)\Gamma_B^{def}\overline \Gamma_{B^\prime}^{abc}\Phi(x_i,k_{i\perp})\Phi^*(x^\prime_i,k^\prime_{i\perp})}{\sqrt{x_1(1-x_1-x_2)x_1^\prime(1-x^\prime_1-x_2)}(k^2-m^{\prime2}_3)}\notag\\
&\times&[(\slashed{p}_3^\prime+m_3^\prime)\gamma^\mu(\slashed k+m_3^\prime)\gamma_\nu(1-\gamma_5)(\slashed p_1+m_1)]_{af}%\notag\\
\times[(\slashed p_1^\prime+m_1^\prime)\gamma^\nu(1-\gamma_5)(\slashed p_3+m_3)]_{cd}[\slashed p_2+m_2]_{be},\notag
\end{eqnarray}
with $k=q+p^\prime_3$, $p_2=p_2^\prime$, 
\begin{eqnarray}
{\mathcal M}^c_T&=&C\int dx_1 dx_2dx^\prime_1 \frac{d^2k_{1\perp}}{2(2\pi)^3}\frac{d^2k_{2\perp}}{2(2\pi)^3}\frac{d^2k^\prime_{1\perp}}{2(2\pi)^3}%\notag\\
 \frac{A_B A_{B'} \epsilon^*_\mu(q)\Gamma_B^{def}\overline \Gamma_{B^\prime}^{abc}\Phi(x_i,k_{i\perp})\Phi^*(x^\prime_i,k^\prime_{i\perp})}{{\sqrt{x_1(1-x_1-x_2)x_1^\prime(1-x^\prime_1-x_2)}(k^2-m_1^2)}}\notag\\
&\times&[(\slashed{p}_3^\prime+m_3^\prime)\gamma_\nu(1-\gamma_5)(\slashed k+m_1)\gamma^\mu(\slashed p_1+m_1)]_{af}%\notag\\
\times[(\slashed p_1^\prime+m_1^\prime)\gamma^\nu(1-\gamma_5)(\slashed p_3+m_3)]_{cd}[\slashed p_2+m_2]_{be},\notag%\\
\end{eqnarray}
with $k=p_1-q$ and $p_2=p_2^\prime$,
\begin{eqnarray}
{\mathcal M}^d_T&=&C\int dx_1 dx_2dx^\prime_1 \frac{d^2k_{1\perp}}{2(2\pi)^3}\frac{d^2k_{2\perp}}{2(2\pi)^3}\frac{d^2k^\prime_{1\perp}}{2(2\pi)^3}%\notag\\
 \frac{A_B A_{B'} \epsilon^*_\mu(q)\Gamma_B^{def}\overline \Gamma_{B^\prime}^{abc}\Phi(x_i,k_{i\perp})\Phi^*(x^\prime_i,k^\prime_{i\perp})}{{\sqrt{x_1(1-x_1-x_2)x_1^\prime(1-x^\prime_1-x_2)}(k^2-m^{2}_3)}}\notag\\
&\times&[(\slashed{p}_3^\prime+m_3^\prime)\gamma_\nu(1-\gamma_5)(\slashed p_1+m_1)]_{af}[\slashed p_2+m_2]_{be}%\notag\\
\times[(\slashed p_1^\prime+m_1^\prime)\gamma^\nu(1-\gamma_5)(\slashed k+m_3)\gamma^\mu(\slashed p_3+m_3)]_{cd}\notag
\end{eqnarray}
with $k=p_3-q$, $p_2=p_2^\prime$,
\begin{eqnarray}
{\mathcal M}^e_T&=&C\int dx_1 dx_2dx^\prime_1 \frac{d^2k_{1\perp}}{2(2\pi)^3}\frac{d^2k_{2\perp}}{2(2\pi)^3}\frac{d^2k^\prime_{1\perp}}{2(2\pi)^3}
 \frac{A_B A_{B'} \epsilon^*_\mu(q)\Gamma_B^{def}\overline \Gamma_{B^\prime}^{abc}\Phi(x_i,k_{i\perp})\Phi^*(x^\prime_i,k^\prime_{i\perp})}{{\sqrt{x_1(1-x_1-x_2)x_1^\prime(1-x^\prime_1-x_2)}(k^2-m^{\prime2}_1)}}\notag\\
&\times&[(\slashed{p}_3^\prime+m_3^\prime)\gamma_\nu(1-\gamma_5)(\slashed p_1+m_1)]_{af}[\slashed p_2+m_2]_{be}%\notag\\
\times[(\slashed p_1^\prime+m_1^\prime)\gamma^\mu(\slashed k+m_1^\prime)\gamma^\nu(1-\gamma_5)(\slashed p_3+m_3)]_{cd},\notag%\\
\end{eqnarray}
with $k=q+p_1^\prime$, $p_2=p_2^\prime$,
\begin{eqnarray}
{\mathcal M}^f_T&=&C\int dx_1 dx^\prime_1 \frac{d^2k_{1\perp}}{2(2\pi)^3}\frac{d^2k^\prime_{1\perp}}{2(2\pi)^3}A_B A_{B'} \epsilon^*_\mu(q)\Gamma_B^{def}\overline \Gamma_{B^\prime}^{abc}
 \frac{1}{\sqrt{x_1(1-x_1-x_2)x_1^\prime(1-x^\prime_1-x^\prime_2)}x_2P^+q^-}\notag\\
&\times&\bigg(-\int dx_2\frac{d^2k_{2\perp}}{2(2\pi)^3}[(\slashed p_2-\slashed q+m_2)\gamma^\mu(\slashed p_2+m_2)]_{be}%\notag\\
%&&\quad
 +\int dx^\prime_2\frac{d^2k^\prime_{2\perp}}{2(2\pi)^3}[(\slashed p^\prime_2+m_2)\gamma^\mu(\slashed p^\prime_2+\slashed q+m_2)]_{be}\bigg)\notag\\
&\times&[(\slashed{p}_3^\prime+m_3^\prime)\gamma_\nu(1-\gamma_5)(\slashed p_1+m_1)]_{af}\Phi(x_i,k_{i\perp})%\notag\\
\times
[(\slashed p_1^\prime+m_1^\prime)\gamma^\nu(1-\gamma_5)(\slashed p_3+m_3)]_{cd}\Phi^*(x^\prime_i,k^\prime_{i\perp}),\label{am}
\end{eqnarray}
with $p^+_2=p_2^{\prime +}$.
And  the coefficient $C$ is defined by
\begin{eqnarray}
C=eV_{us}V^*_{ud} [ z_{1}-z_2+\tau (y_{1}-y_2))],
\end{eqnarray}
%and $p_3=P-p_1-p_2$, $p^\prime_3=P^\prime-p^\prime_1-p^\prime_2$.
The detail of our calculation of ${\mathcal M}^f_T$ in Eq.~\eqref{am} are given in Appendix.A.

For every specific radiative decay process, we need to consider an additional symmetry factor $SF^{T/P}_{B\to B^\prime}$ arising from the flavor and color symmetries of each baryon in the penguin or tree diagram,
\begin{eqnarray}
&&SF^P_{\Sigma^+\to p\gamma}=2,\quad \quad \;\; SF^T_{\Sigma^+\to p\gamma}=4,\quad \quad \;\; %\notag\\
SF^P_{\Lambda/\Sigma^0\to n\gamma}=2,\quad \;\; SF^T_{\Lambda/\Sigma^0\to n\gamma}=2\notag\\
&&SF^P_{\Xi^0\to\Lambda/\Sigma^0\gamma}=2,\quad\;  SF^T_{\Xi^0\to\Lambda/\Sigma^0\gamma}=2,\quad  \;\; %\notag\\
SF^P_{\Xi^-\to \Sigma^- \gamma}=4,\quad \;\;\; SF^T_{\Xi^-\to \Sigma^- \gamma}=0.
\end{eqnarray}
Then one can easily estimate the form factors defined in Eq.~\eqref{ff} by calculating the amplitude shown in %Eq.~\eqref{am} and 
Fig.~\ref{fig1}. The details for numerical analysis will be given in next section.

%%%%%%%%%%%%%%%%%%%%%%%%
\subsection{Numerical results and Phenomenological analysis}
%%%%%%%%%%%%%%%%%%%%%%%%

Before conducting our numerical calculations, one provides the input parameters we used, such as the baryon mass $m_B$, quark mass $m_q$, Fermi-constant $G_F$, electromagnetic coupling constant $\alpha_{em}$, etc. These parameters are listed in Table.~\ref{table2}.

%%%%%%%%%%%%%%%%%%%%%%%%
 \begin{center}
 \begin{table}[!htb]
\caption{ The  input parameters for masses of light baryon octet  and the quark $\overline {\rm MS}$ masses  at  renormalization scale $\mu=1{\rm GeV}$~\cite{ParticleDataGroup:2022pth}.  Here the electromagnetic coupling constant is fixed as $\alpha_{em}=1/137$. The Fermi-constant is $G_F=1.1664\times 10^{-5} \mbox{GeV}^2$. }
\label{table2}%
\begin{tabular}{ccccccccc|ccccc}
\hline \hline
 baryons &\;\;$n$\;\;&\;\;$p$\;\; &\;\;$\Sigma^-$\;\;&\;\;$\Sigma^0$ \;\; &\;\;$\Sigma^+$\;\;&\;\;$\Lambda$\;\; &\;\;$\Xi^-$\;\;&\;\;$\Xi^0$\;\;
 & quark &\;\;$m_u$\;\;&\;\;$m_d$\;\;&\;\;$m_s$\;\;
 \tabularnewline
\hline
mass (${\rm GeV}$) &0.940&0.938&1.198&1.193& 1.189&1.116&1.322&1.315
& mass (MeV) &$2.9 $&$6.3$&$125.6$ 
 \tabularnewline\hline\hline
\end{tabular}
\end{table}
\end{center}
%%%%%%%%%%%%%%%%%%%

The Wilson coefficients $z$ and $y$ in Eq.~\eqref{H} are calculated to the leading order in Ref.~\cite{Bertolini:1994qk,Buchalla:1995vs} as
\begin{eqnarray}
&&z_1=-0.606,\;z_2=1.346,\;y_1=y_2=0,\;
z_{12}=-0.081,\;y_{12}=-0.383.
\end{eqnarray}
The CKM matrix element in  Eq.~\eqref{H} can also be find in Particle Data Group(PDG)~\cite{ParticleDataGroup:2022pth} as
\begin{eqnarray}
&&|V_{us}|=0.224,\;\; |V_{ud}|=0.974,\;
|V_{ts}|=41.5\times 10^{-3},\; |V_{td}|=8.6\times 10^{-3}.
\end{eqnarray}
Then one can calculate that $\tau=-0.00163$.
Since the weak radiative decay is two body decay process, one can easily determine the momentum of final states.
In the rest frame of initial state,  the energy $E$ and absolute value of momentum $|P|$ of radiative photon and final state baryon are obtained as: $|P_{B^\prime}|=E_\gamma=(M_B^2-M_{B^\prime}^2)/2M_B$ and $E_{B^\prime}=(M_B^2+M_{B^\prime}^2)/2M_B$.

Based on the analysis above,  one can extract the form factors by multiplying the structure $\bar u_B\slashed \epsilon u_{B^\prime}$ (or $\bar u_B\slashed \epsilon\gamma_5 u_{B^\prime}$) as shown in Eq.~\eqref{ff}. 
Since the amplitude ${\mathcal M}$ is divide into six parts,  corresponding to the diagrams in Fig.~\ref{fig1}, one can define the amplitude $H^{P/T}_{(5)}$ and $H^i_{(5)}$ as
\begin{eqnarray}
H^{P/T}&=&\sum_{s}\bar u_B\slashed \epsilon u_{B^\prime}{\mathcal M}_{P/T},\;H^i=\sum_{s}\bar u_B\slashed \epsilon u_{B^\prime}{\mathcal M}_{T},\;\;%\notag\\
H^{P/T}_5=\sum_s\bar u_B \slashed \epsilon\gamma_5u_{B^\prime} {\mathcal M}^i_{T},\;H^i_5=\sum_s\bar u_B \slashed \epsilon\gamma_5u_{B^\prime} {\mathcal M}^i_{T},
\end{eqnarray}
where $i=b\sim f$, and the amplitude $H^{T}_{(5)}$ satisfy the relation: $H^{T}_{(5)}=\sum_i H^i_{(5)}$ and $H_{(5)}= H^T_{(5)}+H^P_{(5)}$. For clearly illustrating the specific contribution of each diagram in Fig.~\ref{fig1}, we define the form factor corresponding to each Feynman diagram as
\begin{eqnarray}
a^i=-\frac{H^i M}{6(M+M^\prime)Q_-},\;b^i=\frac{H^i_5 M}{6(M-M^\prime)Q_+}\label{ffH},
\end{eqnarray}
where $i=P,b\sim f$.
For all W exchange Feynman diagrams, one can define the W exchange form factor as
\begin{eqnarray}
a^w&=&\sum_{i} a^i,b^w=\sum_{i} b^i,\; i=b\sim f.
\end{eqnarray}
In our work, we estimate the value of these form factor $a^i(b^i)$ in Table.~\ref{amplitude_ff}. Then the form factor for S-wave $(a)$ and P-wave $(b)$ can be estimated by summing up all the form factor corresponding to each diagram respectively.

%%%%%%%%%%%%%%%%%%%%%%%%
\begin{widetext}
\begin{center}
 \begin{table}[!htb]
\caption{ The corresponding form factor for each Feynman diagram %defined in Eq.~\eqref{ffH}
in unit ($10^{-3}{\rm GeV}^2$). Here we show the numerical results for penguin diagram. Actually, $a^P,b^P$ have  complex phase from CKM matrix as $e^{i\delta}$ with $\delta=-3.14$. The total form factor can be expressed as $a=a^P e^{i\delta}+a^W$ and $b=b^Pe^{i\delta}+b^W$.}
\label{amplitude_ff}%
\begin{tabular}{c|cccccc|cccccc|cccc}
\hline \hline
form factors &$a^P$&$a^b$&$a^c$&$a^d$&$a^e$&$a^f$& $b^P$&$b^b$&$b^c$&$b^d$&$b^e$&$b^f$&$a^W$&$b^W$\tabularnewline
\hline
$\Sigma^+\to p\gamma$&$0.73$&$-5.96$&$-4.77$&$6.75$&$-4.86$&$1.44$&$0.21$&$0.83$&$-0.43$&$-0.43$&$0.49$&$0.54$&$-7.40$&$1.00$\tabularnewline
\hline
$\Lambda\to n\gamma$&$-0.31$&$-0.79$&$-1.58$&$3.97$&$-6.30$&$11.67$&$0.06$&$0.07$&$-0.75$&$-0.62$&$0.34$&$-0.92$&$6.97$&$-1.88$\tabularnewline
\hline
$\Sigma^0\to n\gamma$&$1.06$&$1.44$&$1.97$&$-2.09$&$-1.16$&$-1.68$&$0.30$&$-0.42$&$-0.10$&$-0.29$&$-0.64$&$2.36$&$-1.52$&$0.91$\tabularnewline
\hline
$\Xi^0\to\Lambda\gamma$&$-0.07$&$0.17$&$9.56$&$4.05$&$-2.05$&$-20.52$&$0.006$&$0.32$&$0.42$&$-0.20$&$-1.07$&$-1.62$&$-8.79$&$-2.15$\tabularnewline
\hline
$\Xi^0\to\Sigma^0\gamma$&$0.02$&$11.12$&$-27.56$&$10.49$&$-1.68$&$29.03$&$0.01$&$0.97$&$0.85$&$0.10$&$-0.10$&$-4.98$&$21.40$&$-3.16$\tabularnewline
\hline
$\Xi^-\to \Sigma^- \gamma$&$-0.04$&-&-&-&-&-&$-0.01$&-&-&-&-&-&-&-\tabularnewline
\hline
\end{tabular}
\end{table}
\end{center}
\end{widetext}
%%%%%%%%%%%%%%%%%%%
One can notice that the Feynman diagrams labeled as $(b\sim f)$ have no contribution to the $\Xi^-\to \Sigma^- \gamma$ processes. Therefore we only estimate its $a^P(b^P)$ corresponding to $Q_{12}$ operator. 
However, Table.~\ref{amplitude_ff} shows that the $Q_{12}$ operator makes a very small contribution, and the main contributing operators should be $Q_3\sim Q_6$ for the $\Xi^-\to \Sigma^- \gamma$ process. Here we do not consider the  phenomenological effects from $Q_3\sim Q_6$ induced relevant Feynman diagrams due to the complexity of LFQM. We expect to consider the contributions of $Q_3\sim Q_6$ to these processes in the future.

Correspondingly, we can obtain the branching ratio and asymmetry parameter by the form factors~\cite{Adolph:2022ujd} as 
\begin{eqnarray}
&&Br(B \to B' \gamma)=\frac{G_F^2}{16\pi \Gamma_B}m_B \left(1-\frac{m_{B'}^2}{m_B^2}\right)^3(|a|^2+|b|^2),\;\;\; %\nonumber\\
\alpha=\frac{\Re(a\cdot b)}{|a|^2+|b|^2}\;.\label{br}
\end{eqnarray}
Based on the above form factors, one can predict the  branching ratios and asymmetry parameters in Table.~\ref{pre}.

%%%%%%%%%%%%%%%%%%%%%%%%
 \begin{table}[!htb]
\caption{Our prediction and experimental values for the branching ratios and asymmetries~\cite{BESIII:2022rgl,ParticleDataGroup:2022pth,BESIII:2023fhs}. }
\label{pre}%
\begin{tabular}{c|c|c|c|c}
\hline \hline
\multirow{2}{*}{channel}& \multicolumn{2}{|c|}{branching ratio ($10^{-3}$)}&  \multicolumn{2}{|c}{asymmetry parameter} \cr\cline{2-5}
&pre& exp&pre&exp\tabularnewline
\hline
$\Sigma^+\to p\gamma$&$2.81$&$0.996\pm0.027$&$-0.10$&$-0.652\pm0.060$\tabularnewline
\hline
$\Lambda\to n\gamma$&$3.38$&$0.832\pm0.066$&$-0.25 $&$-0.16\pm0.11$\tabularnewline
\hline
$\Sigma^0\to n\gamma$&$2.79\times 10^{-10}$&-&$-0.22$&-\tabularnewline
\hline
$\Xi^0\to\Lambda\gamma$&$5.54$&$1.17\pm0.07$&$0.23$&$-0.704\pm0.067$\tabularnewline
\hline
$\Xi^0\to\Sigma^0\gamma$&$8.12$&$3.33\pm0.10$&$-0.15$&$-0.69\pm0.06$\tabularnewline
\hline
\end{tabular}
\end{table}

%%%%%%%%%%%%%%%%%%%

We emphasize  that  our prediction has much errors from  the following respects:
\begin{itemize}

\item[*] In Table.~\ref{amplitude_ff}, one can find that the all factors need to combine the contributions of each Feynman diagram so that the errors would be amplified.
%cancel each other out in the form factor $a^i$. Therefore, the error of the total form factor will be enhanced compared to its central value.

\item[*] For every Feynman diagram, the different combinations of $\Gamma^{\alpha\beta\gamma}$ in Eqs.(\ref{wave},\ref{lambda}) lead to much complex structures. This point could enlarge the errors.
%Consequently, the amplitude in Eq.\eqref{am} will have 16 different term contributions. In our calculation, we also find that the contribution of each term cancels out with one another.

\item[*] The contribution of the Feynman diagram (f) is composed of two opposite terms from $p_2$ and $p_2^\prime$ respectively as shown in Eq.~\eqref{am}. 

\item[*] Considering the $20\%$ traditional uncertainty of LFQM, the cancellation of different terms will increase its uncertainty to almost $100\%$.
\end{itemize}

Our predicted central values are consistent with experimental data.
Despite the considerable error in our estimation, our results still possess significant reference value for the experiment.

We found that branching ratio and asymmetry parameter are not sensitive to the polarization.  If further analyzing the polarization  information, one can define forward-backward asymmetry as
\begin{eqnarray}
A^{\pm}_{FB}=\frac{ \left(\int_0^1 d \cos \theta_\gamma \frac{d Br^\pm}{d \cos \theta_\gamma }-\int_{-1}^0 d \cos \theta_\gamma \frac{d Br^\pm}{d \cos \theta_\gamma }\right)}{Br^\pm}.
\end{eqnarray}
Here $\theta_\gamma$ is the angle between the photon momentum and the quantisation axis of the spin in the B-baryon rest frame. $P_B$ and $\lambda_\gamma$ mean the polarization of the initial state baryon and final state photon respectively.  $Br^\pm$ is the abbreviation for branching ratio $Br(B\to B' \gamma)$ with the initial state $B$'s helicity $\lambda_B=\pm \frac{1}{2}$. The polarization branching ratio $Br^\pm$ can be expressed by form factors as
\begin{eqnarray}
\frac{d Br^\pm(B \to B' \gamma)}{d\cos(\theta_\gamma)}&=&\frac{G_F^2}{16\pi \Gamma_B}m_B \left(1-\frac{m_{B'}^2}{m_B^2}\right)^3%\notag \\
 \times(|a|^2+|b|^2\mp 2Re(a \cdot b^\ast)\cos(\theta_\gamma)).
\end{eqnarray}
In this case, the forward-backward asymmetry  can be simplified as $A^{\pm}_{FB}=\mp Re(a \cdot b^*)/(|a|^2+|b|^2)$.
Therefore, we predict $A^{\pm}_{FB}$ as
\begin{eqnarray}
&&A^{\pm}_{FB}(\Sigma^+\to p\gamma)=\mp0.096,\;\; A^{\pm}_{FB}(\Lambda\to n\gamma)=\mp0.249,\;\; %\notag\\
A^{\pm}_{FB}(\Sigma^0\to n\gamma)=\mp0.221,\notag\\
&&A^{\pm}_{FB}(\Xi^0\to\Lambda\gamma)=\pm0.233,\;\; 
A^{\pm}_{FB}(\Xi^0\to\Sigma^0\gamma)=\mp 0.145.
\end{eqnarray}

%%%%%%%%%%%%%%%%%%%%%%%%
\subsection{Discussion in certain scenarios}
%%%%%%%%%%%%%%%%%%%%%%%%

Through our calculations, we have identified intriguing patterns. First, the contribution of the penguin Feynman diagram is significantly smaller than that of the W exchange Feynman diagram. 
Secondly, the value of the parity-conserving form factor $a$ is on the order of $\sim 1$, which is greater than the parity-violating one, $b \sim 0.1$.

Despite the significant errors in the form factors, the orders of  form factors remain accurate.
In the definition of the asymmetry parameter in Eq.~\eqref{br}, it is evident that the asymmetry parameter is primarily determined by the disparity between the two form factors $a$ and $b$.
Consequently, the predictive error of the asymmetry parameter is expected to decrease significantly. Furthermore, the disparity in $\alpha(\Sigma^+\to p\gamma)$ indicates that these processes may entail a substantial strong phase, defined as 
 \begin{eqnarray}
a&=&|a^P|e^{i\delta}e^{i\delta^P_a}+|a^W|e^{i\delta^W_a},\;\; 
b=|b^P|e^{i\delta}e^{i\delta^P_b}+|b^W|e^{i\delta^W_b},
\end{eqnarray}
where we assume that the W exchange Feynman diagram have a global strong phase $\delta^W_a(\delta^W_b)$.

Unfortunately, our model cannot predict the strong phase. The values of strong phase could be determined by the experimental data via the following assumptions:
\begin{itemize}
\item[*] We assume accuracy of our results even if neglecting the errors in our calculation.
\item[*] The different processes have the same strong phases.
\item[*] We ignore  the strong phase  of penguin Feynman diagram $\delta^P_{a,b}$ due to small contributions.
\end{itemize}
In this case, we can perform a rough global fit for the two strong phases $\delta^W_{a,b}$. Utilizing the non-linear least-$\chi^2$ method~\cite{lsq} and the form factors in Table~\ref{amplitude_ff}, we present the fit results in Table~\ref{gf}.

%%%%%%%%%%%%%%%%%%%%%%%%
 \begin{table}[!htb]
\caption{The global analysis of strong phase $\delta^W_a(\delta^W_b)$ in the above scenarios.  }
\label{gf}%
\begin{tabular}{c|c|c}
\hline \hline
channel& Br ($10^{-3}$)&  $\alpha$\tabularnewline
\hline
$\Sigma^+\to p\gamma$&$1.928\pm0.015$&$-0.124\pm0.078$\tabularnewline
\hline
$\Lambda\to n\gamma$&$2.8465\pm0.0064$&$-0.16\pm0.14$\tabularnewline
\hline
$\Xi^0\to\Lambda\gamma$&$5.7069\pm0.0020$&$0.15\pm0.12$\tabularnewline
\hline
$\Xi^0\to\Sigma^0\gamma$&$8.14462\pm0.00080$&$-0.092\pm0.073$\tabularnewline
\hline
 $\chi^2/d.o.f=3.80$&$\delta^W_a=-3.02\pm0.77$&$\delta^W_b=2.29\pm0.79$\tabularnewline
\hline
\end{tabular}
\end{table}
%%%%%%%%%%%%%%%%%%%

In our fit, we used the central values of the experimental data as input, and we assigned a $100\%$ error to account for the uncertainty in our input form factors.

Using the fitted strong phases, we can predict the CP violation observable $A_{CP}$, as 
\begin{eqnarray}
&&A_{CP}(\Sigma^+\to p\gamma)=1.91\times 10^{-5},\quad A_{CP}(\Lambda\to n\gamma)=2.55\times 10^{-6},\quad A_{CP}(\Sigma^0\to n\gamma)=-5.16\times 10^{-4},\notag\\
&&A_{CP}(\Xi^0\to\Lambda\gamma)=8.81\times 10^{-7},\quad A_{CP}(\Xi^0\to\Sigma^0\gamma)=2.24\times 10^{-7}.
\end{eqnarray}
Here we only give the center value due to  large errors in the form factor.

One can observe that all the processes we consider exhibit very small CP violation. However, it must be mentioned that the global analysis is a preliminary and rough analysis, and it can only be applied within the specific scenarios listed in this section. Therefore, a more complete and comprehensive analysis will require a more precise theoretical calculation and additional experimental data in the future.

%%%%%%%%%%%%%%%%%%%%%%%%
\section{Summary}
%%%%%%%%%%%%%%%%%%%%%%%%
In our study, we explore the LFQM in a three-quark picture for the light baryon octet. Considering $SU(3)_f$ flavor symmetry, we construct symmetry wave functions to  express the baryon state using three constituent quarks. One  unique advantage of this model is that it allows us to estimate the Feynman diagram involving the exchange of the W boson between two constituent quarks. Therefore, we investigate a class of hyperon weak radiative decays induced by $s\to d$. In our analysis, we neglect the contributions of QCD penguin Hamiltonians $Q_3\sim Q_6$ due to their small Wilson coefficients.
Our calculations indicate that the main contribution to hyperon weak radiative decays induced by $s\to d$ comes from the W exchange Feynman diagram.

By extracting the parity-conserving and parity-violating form factors $a$ and $b$ from Table.~\ref{amplitude_ff}, we can estimate many phenomenological observables, including branching ratios, asymmetry parameters, and forward-backward asymmetries. Our results indicate that the forward-backward asymmetries for $\Sigma^+\to p\gamma$, $\Lambda\to n\gamma$, $\Sigma^0\to n\gamma$, and $\Xi^0\to\Sigma^0\gamma$ are on the order of $10^{-3}$ and have a high likelihood of being measured.

At the end of our work, we also provide an integrated discussion with some scenarios. Utilizing the scenarios listed in Sec.~IIIC, we can perform a preliminary rough global analysis and predict the CP violation observable $A_{CP}$. While our analysis suggests that the $A_{CP}$ of hyperon weak radiative decays is near zero, it must be noted that a more complete and comprehensive analysis, which requires more precise theoretical calculations and additional experimental data in the future without assumptions, may result in an increase in the value of $A_{CP}$.

%%%%%%%%%%%%%%%%%%%%%%%%
\section*{Acknowledgements}
%%%%%%%%%%%%%%%%%%%%%%%%

%This work was supported by National Natural Science Foundation of China under grant No.U2032102, 12090064, 12125503, by Natural Science Foundation of Shanghai and
This work was supported by IBS under the project code, IBS-R018-D1. The work of Yu Ji Shi is supported by Opening Foundation of Shanghai Key Laboratory of Particle Physics and Cosmology under Grant No.22DZ2229013-2 and National Natural Science Foundation of China under Grant No.12305103. The work of Zhen Xing Zhao was supported in part by the National Natural Science Foundation of China under Grant No.12065020.
	\begin{appendix}
 \section{The simplify of amplitude of  Feynman diagram (f)}
 After expanding the initial and final baryon states by LFQM, the Fig. f amplitude  is expressed as
\begin{eqnarray}
{\mathcal M}^f_T&=&C \sum^3_{i=1}\int d^4 x \{d\tilde {p_i}\}\{d\tilde {p_i^\prime}\}(2(2\pi)^3)^2 e^{-i(p_2-p_2^\prime-q)x}%\notag\\
\times \delta^3(\bar P-p_1-p_2-p_3) \delta^3(\bar P^\prime-p^\prime_1-p^\prime_2-p^\prime_3)\notag\\
&\times&\frac{A_B A_{B'}\epsilon^*_\mu(q)  \Phi(x_i,k_{i\perp})\Phi^*(x^\prime_i,k^\prime_{i\perp})}{\sqrt{p^+_1p^+_2p_3^+P^+p^{\prime+}_1p^{\prime+}_2p_3^{\prime+}P^{\prime+}}}%\notag\\
\times\Gamma_B^{def}\overline \Gamma_{B^\prime}^{abc} [(\slashed {p}^\prime_2+m_2^\prime)\gamma_\nu(\slashed p_2+m_2)]_{be} \notag\\
&\times&[(\slashed p^\prime_3+m_3^\prime)\gamma_\mu(1-\gamma_5)(\slashed p_1+m_1)]_{af}
\times [(\slashed p^\prime_1+m^\prime_1)\gamma^\mu(1-\gamma_5)(\slashed p_3+m_3)]_{cd}.\label{fd}
\end{eqnarray}
Before integrating out the momentum $p_2$ and $p_2^\prime$, one can prove the equation from the contour integration as
\begin{eqnarray}
\int \frac{d^4 p}{(2\pi)^4}e^{ipx}\frac{i(\slashed p+m)}{p^2-m^2+i\epsilon}&=&\int\frac{d^3p}{2(2\pi)^3}e^{ipx}\frac{\slashed p+m}{2p^+}\bigg|_{x^+<0},\notag\\
\int \frac{d^4 p}{(2\pi)^4}e^{-ipx}\frac{i(\slashed p+m)}{p^2-m^2+i\epsilon}&=&\int\frac{d^3p}{2(2\pi)^3}e^{-ipx}\frac{\slashed p+m}{2p^+}\bigg|_{x^+>0}.\notag\\ \label{ci}
\end{eqnarray}
Then we consider the  $p_2$ and $p_2^\prime$ part in Eq.~\eqref{fd} as
\begin{eqnarray}
I&=&\int d^4x\int \frac{d^3p_2}{2(2\pi)^3}\frac{d^3p^\prime_2}{2(2\pi)^3}\frac{e^{-i(p_2-p_2^\prime-q)x}}{\sqrt{p_2^+p_2^{\prime +}}}%\notag\\
\times [(\slashed p^\prime_2+m_2)\gamma_\nu(\slashed p_2+m_2)]_{be}.
\end{eqnarray}
 For our analysis, we obtain $p^+_2=p_2^{\prime +}$ by
setting $q^+=q_\perp=0$ and $q^-\neq 0$. In this case, $I$ can be expressed by
\begin{eqnarray}
I&=&\int d^4x\int \frac{d^3p_2}{2(2\pi)^3}\frac{d^3p^\prime_2}{2(2\pi)^3}\frac{e^{-i(p_2-q)x }}{p_2^{\prime +}}%\notag\\
 e^{ip_2^\prime x}[(\slashed p^\prime_2+m_2)\gamma_\nu(\slashed p_2+m_2)]_{be}\bigg|_{x^+<0}\notag\\
&+&\int d^4x\int \frac{d^3p_2}{2(2\pi)^3}\frac{d^3p^\prime_2}{2(2\pi)^3}\frac{e^{i(p_2^\prime +q)x}}{p_2^+}%\notag\\
 e^{-ip_2 x}[(\slashed p^\prime_2+m_2)\gamma_\nu(\slashed p_2+m_2)]_{be}\bigg|_{x^+>0}.
\end{eqnarray}
Using the formula in Eq.~\eqref{ci}, one can derive that
\begin{eqnarray}
I&=&2\int d^4x \int \frac{d^3p_2}{2(2\pi)^3}\frac{d^4p^\prime_2}{(2\pi)^4}\frac{ie^{-i(p_2-p_2^\prime-q)x}}{p_2^{\prime 2}-m_{2}^2}%\notag\\
 \times [(\slashed p^\prime_2+m_{2})\gamma_\nu(\slashed p_2+m_2)]_{be}\notag\\
&+&2\int d^4x \int \frac{d^3p^\prime_2}{2(2\pi)^3}\frac{d^4p_2}{(2\pi)^4}\frac{ie^{-i(p_2-p_2^\prime-q)x}}{p_2^{2}-m_2^2}%\notag\\
 \times [(\slashed p^\prime_2+m_2)\gamma_\nu(\slashed p_2+m_2)]_{be}\notag\\
&=&2i\bigg(\int \frac{d^3p_2}{2(2\pi)^3} \frac{[(\slashed p_2-\slashed q+m_2)\gamma_\nu(\slashed p_2+m_2)]_{be}}{(p_2-q)^2-m_{2}^2} %\notag\\
%&& 
+\int \frac{d^3p^\prime_2}{2(2\pi)^3} \frac{[(\slashed p^\prime_2+m_2)\gamma_\nu(\slashed p_2+\slashed q+m_2)]_{be}}{(p^\prime_2+q)^2-m_2^2} \bigg).
\end{eqnarray}
Here we use $\int d^4xe^{-ix(p_2-p_2^\prime-q)}=(2\pi)^4\delta^4(p_2^\prime -p_2-q)$ and derive $p_2^\prime =p_2+q$ after integrating over $p_2$ or $p_2^\prime$.
By substituting the integration $I$ into Eq.~\eqref{fd},  one can derive the amplitude in Eq.~\eqref{am} correspondingly .
 \end{appendix}

\end{document}